\documentclass{iopart}

\usepackage{iopams,setstack}
\usepackage{epsf}
\usepackage{cite}
\usepackage{enumerate}

\newcommand{\bd}{\begin{displaymath}}
\newcommand{\ed}{\end{displaymath}}
\newcommand{\be}{\begin{equation}}
\newcommand{\ee}{\end{equation}}
\newcommand{\ba}{\begin{eqnarray}}
\newcommand{\ea}{\end{eqnarray}}

\begin{document}



\paper[Coherence loss and revivals in atomic interferometry]
{Coherence loss and revivals in atomic interferometry:
A quantum--recoil analysis}

\author{M Davidovi\'c$^1$, A S Sanz$^2$, M Bo\v zi\'c$^3$ and
D Arsenovi\'c$^3$}

\address{$^1$Faculty of Civil Engineering, University of Belgrade,
Bulevar Kralja Aleksandra 73, 11000 Belgrade, Serbia}

\address{$^2$Instituto de F\'{\i}sica Fundamental (IFF--CSIC),
Serrano 123, 28006 Madrid, Spain}

\address{$^3$Institute of Physics, University of Belgrade,
Pregrevica 118, 11080 Belgrade, Serbia}

\eads{\mailto{asanz@iff.csic.es}}

\begin{abstract}
The coherence effects induced by external photons coupled to matter
waves inside a Mach-Zehnder three-grating interferometer are analyzed.
Alternatively to atom-photon entanglement scenarios, the
model considered here only relies on the atomic wave function and
the momentum shift induced in it by the photon scattering events. A
functional dependence is thus found between the observables, namely
the fringe visibility and the phase shift, and the transversal
momentum transfer distribution. A good quantitative agreement is
found when comparing the results obtained from our model with the
experimental data.
\end{abstract}







\section{Introduction}
 \label{sec1}

The remarkable refinement reached in matter wave interferometry in the
last decades \cite{rauch,cronin1} has made possible to explore
experimentally fundamental key questions about wave particle duality
and complementarity that have been studied since the very inception
of quantum mechanics \cite{bohr,debroglie}.
In this regard, Chapman \etal~\cite{chapman} carried
out an outstanding experiment in 1995, where the influence of
photon-atom scattering events (inside an atomic Mach-Zehnder
interferometer) on the coherence properties of an atomic beam was
investigated. This experiment was interpreted as a realization with
atoms of Feynman's ``which-way'' {\it gedankenexperiment} \cite{feynman}.

The most intriguing result from Chapman's experiment was the revival
of fringe contrast beyond the limits predicted by the complementary
principle \cite{chapman,cronin1,schmiedmayer}.
Furthermore, it was also observed \cite{chapman} a regain of
fringe contrast after post-selecting atoms at the exit of the
interferometer according to the momentum transferred in the
photon-atom scattering process. The regain of interference due to
post-selection in momentum space had been previously reported for
optical \cite{mandel3} and neutron \cite{rauch2}
experiments with presence of resonant
spin-flipper fields.
In the case of the neutron experiments, a
spectral modulation effect was observed by means of a proper
post-selection procedure, where the spatial shift of the wave trains
greatly exceeds the coherence length of the neutron beams traversing
the interferometer \cite{rauch,rauch2}.

By the time when the paper by Chapman \etal \cite{chapman} was
published, a controversy on the origin of the disappearance of
interference in ``which-way'' (actually, ``which-slit'') double-slit
experiments was already in fashion: recoil {\it vs} decoherence.
At a first glance, it seems that the primacy of recoil arguments
\cite{wiseman1} has been contested in favor of more general decoherence
mechanisms, based on considering the entanglement between the observed
system and its environment to be the source of the system loss of
fringe contrast or visibility.
Nevertheless, Storey \etal \cite{storey1} argued that,
whenever interference is destroyed, transverse momentum has to be
transferred according to the uncertainty principle.

Revivals observed beyond the limit of the complementarity principle
enforced Chapman \etal \cite{chapman} and Cronin \etal \cite{cronin1}
to argue that ``the momentum recoil by itself can not explain the loss
of contrast (as it can in the diffraction experiments), but the path
separation at the point of scattering and the phase shift imprinted
by the entanglement in the scattering process must also be taken into
account''.
In addition, Cronin \etal \cite{cronin1} argued that ``focusing on the
which-way information carried away by the scattered photons is not
the only way decoherence may be understood. An alternative, but
completely equivalent picture involves the phase shift between the
two components of the atomic wave function''.
These two views (which-way and dephasing) ``correspond to two different
ways to describe the scattered photon (position basis versus momentum
basis).
In these two cases, an observer in the environment can determine
either which path the atom took or else the phase shift of its
fringe pattern. The key point is that when the experimenter is
completely ignorant of the state of the scattered photons, whether
an apparatus has been set up to measure them or not, the which-path
and phase diffusion pictures are equally valid (Stern {\it et al}.,
1990, \cite{stern}). Both predict decoherence, i.e., loss of contrast''
\cite{cronin1}.

It is important to note that the apparatus of Chapman \etal
\cite{chapman} was set up to detect atoms, but not to measure
the state of the scattered photons.
Because of this, in the present work we study this
experiment using a model \cite{arsenovic1,arsenovic2}
that focuses on atomic states. It
accounts for the effects caused on the atom time-dependent wave
function by the interferometer as well as the (environmental)
photons scattered from the atoms when the latter are excited in a
resonance fluorescence state by a laser beam. Due to the negligible
time-scales involved in the dynamics of the atom-photon scattering
process (i.e., the absorption and then re-emission of the photon by
the atom) compared with the time-scales involved in the experiment,
the photon atom resonance scattering is described as a sudden change
of the atom wave function accompanying the momentum transfer between
the photon and the atom. Hence we assume each atom can be
individually described by a pure state, and only when a collection
of atoms is considered statistically, the decoherence effect arising
from the photon-induced momentum displacements becomes apparent.
More specifically, here we use the probability distribution of
transverse momentum transfer to an atom in resonance fluorescence
derived by Mandel \cite{mandel1,mandel2}
from the angular distribution of
spontaneously emitted photons.

According to such a model, here we present a functional dependence
between the experimental observables, namely the fringe visibility
and the phase shift, and the statistical distribution of photon-atom
transversally transferred momentum. From this relationship, a direct
connection is established between the coherence losses and
subsequent revivals undergone by the atoms, which arise as a
consequence of the statistical distribution of the sudden momentum
shifts induced in the atomic wave function by the photons
(scattering-mediated momentum transfer processes). Furthermore, when
some particular choices of momentum transfers are considered by
selecting the outgoing atoms according to some prescribed momentum
distributions, i.e., by post-selecting the atoms, a regain of the
coherence is observed.
As it is shown, these results are in good agreement (both qualitatively
and also quantitatively) when compared with the experimental data
reported by Chapman \etal \cite{chapman}.
Note therefore that this simple model thus provides a self-consistent
explanation of the experiment based on first-principle-like arguments
rather than only a best fitting to some suitable function.

This work is organized as follows. In Section \ref{sec2}, to be
self-contained, we start by briefly introducing the experimental
setup used by Chapman \etal \cite{chapman} as well as an also brief
description of the two types of experiments they carried out. In
Section \ref{sec3}, we introduce our theoretical description of this
experiment together with the analytical tools that arise from it to
later on evaluate the fringe visibility and phase shift, which are
compared with the experimental data. As it will be seen, this
entails the two features of a quantum particle within the same
experiment: wave and corpuscle. In other words, with each individual
atom that enters into and passes through the three-grating
Mach-Zehnder interferometer, and then arrives at the detector, there
is a wave associated, which is described by a coherent wave function
or pure state. In Section \ref{sec4}, results for different functional forms
of the transversal momentum transfer distribution are analyzed and
discussed. As it is shown, when these results are directly compared
with the experimental data reported in \cite{chapman}, a good agreement
is found even without using any best-fit method, but just introducing
the experimental parameters into the functional forms derived from
our theoretical model. Finally, the main conclusions arising from
this work have been summarized in Section \ref{sec5}.


\section{Description of the experiment}
 \label{sec2}

\begin{figure}
 \begin{center}
 \epsfxsize=13cm {\epsfbox{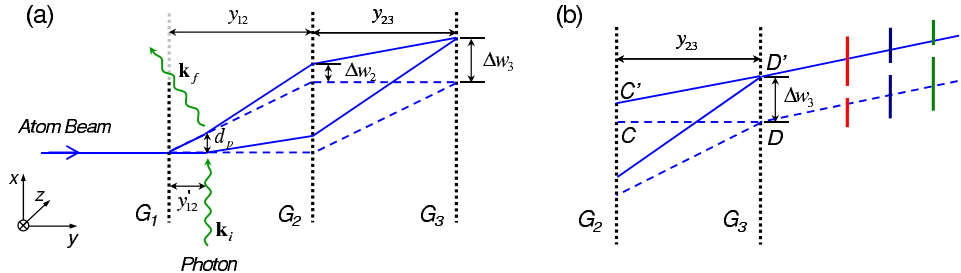}}
 \caption{\label{fig1}
  (a) Scheme of the experimental setup used by Chapman \etal
  \cite{chapman} to conduct their experiments on atom interferometry.
  Essentially, it consists of a Mach-Zehnder three-grating
  interferometer, where atoms are acted by external photons between
  the first and second gratings ($G_1$ and $G_2$).
  (b) Scheme showing the postselection slits behind the third grating
  G3; each one gives rise to a different postselection momentum
  transfer distribution (see Section~\ref{sec3}).}
 \end{center}
\end{figure}

In the experimental setup utilized by Chapman \etal \cite{chapman} (a
sketch is shown in figure~\ref{fig1}a), a beam of atomic sodium with a
narrow velocity distribution is produced, collimated and launched
through an atomic Mach-Zehnder interferometer.
The interferometer consists of
three 200~nm period nanofabricated Ronchi diffraction gratings
(indicated by the vertical dotted lines in figure~\ref{fig1}a)
separated by $L = 65$~cm. Each grating acts as a coherent beam
splitter \cite{bozic1}, with the zeroth and first order maxima being
the relevant ones.

A polarized laser beam behind the first grating, $G_1$, is switched on
with the direction of the beam being parallel to this slit. This
laser leads the atoms to a resonant excited state, from which they
decay back to the ground state via spontaneous emission. The atomic
flux collected behind the third grating, $G_3$ (see figure~\ref{fig1}a),
was then
measured as a function of a shift $\Delta x_3$
produced in this grating along
the $x$-axis, with the laser both off and on. This measurement was
performed considering different values of the distance $y'_{12}$
between $G_1$ and the laser beam. Then, next, the same set of
measurements was
repeated, but adding a selection slit behind $G_3$, in front of the
detector (see figure~\ref{fig1}b). Each selection slit was associated with a
particular range of values of the transferred transverse momentum.

The dependence of the measured values of the number of detected
atoms on the shift $\Delta x_3$, given by
\be
 N(\Delta x_3) = \bar{N}  \left[ 1 + C \cos \left(
  \frac{2\pi}{d_g}\ \Delta x_3 + \varphi \right) \right] ,
 \label{number}
\ee
revealed interference \cite{chapman}.
In this expression, $\bar{N}$ is the average atom
count rate, $d_g$ is the period of the grating and $C$ is the relative
contrast (or fringe visibility).
When the laser was off, the contrast $C$ was typically about 20\% and
the phase $\varphi$ was zero. When the laser was turned on, photon
scattering events before and immediately after $G_1$ does not affect
either the contrast $C$ or the phase. However, as $y'_{12}$ increases,
the contrast decreases, first linearly and then it sharply falls to
zero. Afterward few revivals were observed. This behavior can be
seen in figure~2 of \cite{chapman},
where the relative contrast (visibility)
was represented as a function of $d_p/\lambda_i$, with $\lambda_i$
being the photon wavelength and
\be
 d_p = \left( \frac{2\pi}{kd_g} \right) y'_{12} .
 \label{dp}
\ee
Chapman \etal \cite{chapman} interpreted the quantity $d_p$ as
``the relative
displacement of the two arms of the interferometer at the point of
scattering''. However, Bo{\v z}i{\' c} \etal \cite{arsenovic2} pointed
out that this
quantity is equal to the separation between the two paths associated
with the zeroth and first order interference maxima only in the far
field, behind $G_1$. On the contrary, in the near field, $d_p$
is equal to
the distance between the prolongations of such paths. This
distinction should be taken into account when interpreting the
experimental data, since the photon-atom scattering events in this
experiment take place in the near field. In this work, this is
explained in detail, taking into account the following fact:
\be
 y'_{12} = \frac{kd_g}{2\pi}\ d_p = \frac{d_p}{\lambda_i}
  \frac{kd_g}{k_i} = \frac{d_p}{\lambda_i} \frac{L_T}{2}
  \frac{\lambda_i}{d_g} ,
 \label{dist}
\ee
where $L_T = 2d_g^2/\lambda$ is the so-called Talbot distance
\cite{sanz-talbot}.
In the experiment, the ratio $d_p/\lambda_i$ ranges between 0 and 2.
From the values of the other experimental parameters, it follows that
$y'_{12} \in [0, 19.09]$~mm and the Talbot distance is $L_T = 6.48$~mm.

The same set of measurements was repeated, but this time adding a
selection slit behind $G_3$, in front of the detector (see
figure~\ref{fig1}b).
More specifically, this was done by arranging slits in three
different positions, each selection slit being associated with a
particular range of values of the transverse momentum $\Delta k_x$
transferred
to the atom (i.e., with a particular momentum transfer
distribution).  This was possible because the deflection of the atom
at the third grating, $\Delta w_3$, is proportional to $\Delta k_x$,
the transverse momentum transferred to the atom.
The curves shown in figure~3 of \cite{chapman}
show a substantial regain of contrast over the whole range of
values for $d_p/\lambda_i$.
In particular, a 60\% of the contrast lost at $d_p \approx \lambda_i/2$
was regained.

From these results, Chapman \etal \cite{chapman}
concluded that the decrease of
contrast to zero in the range $0 < d_p/\lambda_i < 0.5$
confirms the complementarity in
quantum mechanics, which suggests that fringe contrast must
disappear when it is possible to acquire which-way information,
i.e., for $d_p/\lambda_i > 0.5$.
Consequently, one should expect that beyond this
value no coherence should be possible. On the contrary, the
experiment revealed that the atomic coherence displayed revivals in
the relative contrast beyond the first zero, thus allowing the atoms
to also display some wave-like behavior beyond the limits of
complementarity. Furthermore, in the second part of the experiment,
it was also observed that the coherence could be regained; actually,
no zero values were observed in the relative contrast.

In our opinion, analyzing this kind of experiments in terms of the
idea of complementarity might result confusing, though very
widespread. This was already pointed out by Englert \cite{englert}
in 1996,
who warned about the misunderstandings that may arise from the use
of concepts like wave-particle duality unless they are clearly
specified and disambiguated. As it is shown below, in the model here
described, such concepts, namely wave and particle, are not mutually
exclusive, but they both coexist in the experiment, giving a good
account of the experimental data. In particular, the wave aspect of
the atom is kept all the way through the interferometer, the photon
only causing a deviation of its translational motion (due to the
kick and subsequent momentum transfer during the scattering event).

Having in mind these ideas and the scheme displayed in
figure~\ref{fig1}a, in the derivations presented below, we assume
the atomic beam incident onto the grating $G_1$ (at $y = 0$) can be
well approximated by a monochromatic or plane wave of finite
transverse width with wavelength $\lambda$ and wave vector ${\bf k}
= (2\pi/\lambda) \hat{\bf y}$. If the atomic beam cross-section is
also assumed to be wide enough (in the experiment, this
cross-section is about two orders of magnitude larger than the
grating period \cite{turchette}), not only it will cover a
relatively large number of slits, but also an important extension
along the $z$-direction. This causes a symmetry along the
$z$-direction, which allows us to simplify the analysis by reducing
it to the $XY$-plane (for fixed $z$, e.g., $z=0$).


\section{Theoretical approach}
 \label{sec3}


\subsection{Atom's wave function evolution accompanying atom's passage
through the interferometer}
\label{sec31}

Taking into account the description of the experiment made above,
now we are going to analyze it here according to our model. Thus,
consider the incident atomic wave function associated with atoms
having a velocity $v$ is given by
\be
 \Psi_{\rm inc}(x,y,t) = e^{-i\omega t} e^{iky} \psi_{\rm inc} (x) ,
 \label{e23}
\ee
where $\hbar\omega = \hbar^2 k^2/2m$, $v = \hbar k/m$ and
$\psi_{\rm inc}(x)$ describes the width of the initial wave function
along the transverse direction. In the paraxial approximation, the
outgoing wave evolving freely after the diffraction caused by $G_1$
is approximated by
\be
 \Psi(x,y,t) = e^{-i\omega t} e^{iky} \psi^{\rm tr} (x,t) .
\ee
This function is a product of the plane wave along the longitudinal
$y$-direction by the ``transverse'' wave function
\be
\fl
 \psi^{\rm tr}(x,t) = \frac{1}{\sqrt{2\pi}}\
  \int_{-\infty}^\infty c(k_x) e^{ik_x x - i\hbar k_x^2 t/2m} dk_x
  = \frac{1}{\sqrt{2\pi}}\
  \int_{-\infty}^\infty C(k_x,t) e^{ik_x x} dk_x .
 \label{e20b}
\ee
which describes the evolution along the $x$-direction. The function
$c(k_x)$ is the Fourier transform of the function $\psi^{\rm tr}(x,0)$
which is determined by $\psi_{\rm inc}(x)$ through the relation
\be
 \psi^{\rm tr}(x,0) = T(x) \psi_{\rm inc} (x) ,
 \label{e24}
\ee
where $T(x)$ is the given transmission function of the grating $G_1$
located at $y=0$.
It is also the transmission function of grating $G_2$.
More explicitly,
\ba
 c(k_x) & = & \frac{1}{\sqrt{2\pi}}\
  \int_{-\infty}^\infty T(x) \psi_{\rm inc}(x) e^{-ik_x x} dx ,
 \label{e16} \\
 C(k_x,t) & = & \frac{1}{\sqrt{2\pi}}\
  \int_{-\infty}^\infty \psi^{\rm tr}(x,t) e^{-ik_x x} dx
  = c(k_x) e^{ik_x^2\hbar t/2m} ,
 \label{e16b}
\ea
Evidently, $C(k_x,t)$ is the time-dependent transverse wave function
in momentum representation.

Taking into account the length scales involved in the experiment, the
paraxial approximation can be considered a good approximation.
This implies, first, that the particle motion parallel to the
$y$-direction can be treated as a quasi-classical (uniform) motion,
i.e., satisfying the relation $y = vt$, with $v = \hbar k/m =
2\pi\hbar/\lambda$.
Second, the wave function (\ref{e24}) behind the grating $G_1$ is such
that $c(k_x)$ is relevant
only for $k_x^2 \ll k_y^2 \approx k^2 = k_x^2 + k_y^2$ (in other words,
the spreading of the wave function is much slower than its propagation
along the $y$-direction \cite{sanz-JPA}).
Accordingly, equation~(\ref{e20b}) can be parameterized in terms
of the $y$-coordinate or, equivalently, the (propagation) time~$t$.

In the passage from $G_2$ to $G_3$ as well as beyond $G_3$, a similar
analysis can be conducted (see below). However, at a time $t'_{12}$
and a distance $y'_{12} = v t'_{12} = (\hbar k/m) t'_{12}$
after the grating $G_1$ the atom absorbs and re-emits a
photon. This process induces a sudden change $\Delta k_x$ in the atomic
transverse momentum which is accompanied by the sudden change of the
evolution of atom's wave function. Arsenovi\'c \etal \cite{arsenovic1}
determined
the evolution of atom's wave function after photon atom scattering
by assuming that atom's wave function in momentum representation
after photon atom scattering $C_{\Delta k_x}(k_x,t)$ has to satisfy:
\be
 |C_{\Delta k_x}(k_x,t'_{12})|^2 =
  |C(k_x - \Delta k_x,t'_{12})|^2 .
 \label{cond1}
\ee
The corresponding transverse wave function at time $t'_{12}$, in
accordance to (\ref{e20b}) is then given by
\be
 \psi_{\Delta k_x}^{\rm tr}(x,t'_{12}) = \frac{1}{\sqrt{2\pi}}\
  \int_{-\infty}^\infty C_{\Delta k_x}(k_x,t'_{12}) e^{ik_x x} dk_x .
 \label{e20}
\ee
It should satisfy
\be
 |\psi_{\Delta k_x}^{\rm tr}(x,t'_{12})|^2 =
  |\psi^{\rm tr}(x,t'_{12})|^2 .
 \label{cond2}
\ee
As shown by Arsenovi\'c \etal \cite{arsenovic1}, from equations
(\ref{e16b})-(\ref{e20}) it  follows that condition (\ref{cond2})
will be fulfilled if
\be
 C_{\Delta k_x}(k_x,t'_{12}) = C(k_x - \Delta k_x,t'_{12}) .
 \label{cond3}
\ee
Substituting (\ref{cond3}) into (\ref{e20}) and then using (\ref{e16b}),
one finds that just after the photon-atom scattering event, the atomic
wave function becomes
\be
\fl
 \psi_{\Delta k_x}^{\rm tr} (x,t'_{12}) = \frac{1}{\sqrt{2\pi}}\
  e^{-i\Delta k_x^2 \hbar t'_{12}/2m}
  \int_{-\infty}^{\infty}
  c(k_x - \Delta k_x) e^{-ik_x^2 \hbar t'_{12}/2m
   + i k_x (x + \Delta x_0)} dk_x ,
 \label{e28}
\ee
where
\be
 \Delta x_0 = \frac{\Delta k_x \hbar t'_{12}}{m}
  = \left(\frac{\Delta k_x}{k}\right) y'_{12} .
 \label{e29}
\ee
Assuming (\ref{e28}) keeps the same form at any $t>t'_{12}$, we may
write:
\be
\fl
 \psi_{\Delta k_x}^{\rm tr} (x,t) = \frac{1}{\sqrt{2\pi}}\
  e^{-i\Delta k_x^2 \hbar t/2m}
  \int_{-\infty}^{\infty}
  c(k_x - \Delta k_x) e^{-ik_x^2 \hbar t/2m
   + i k_x (x + \Delta x_0)} dk_x .
 \label{e30}
\ee
By changing now the integration variable $k'_x = k_x - \Delta k_x$,
(\ref{e30}) transforms into
\be
\fl
 \psi_{\Delta k_x}(x,t)
 = e^{i\Delta k_x(x + \Delta x_0) - i\Delta k_x^2 \hbar t/m}
  \int_{-\infty}^{\infty} c(k'_x) e^{-i{k'}_x^2 \hbar t/2m}
     e^{ik'_x (x + \Delta x_0 - \hbar t \Delta k_x/m)} dk'_x .
 \label{e32}
\ee
This wave function describes the evolution of (\ref{e20b}) after the
scattering event (i.e., for $t>t'_{12}$ or, equivalently, $y>y'_{12}
=(\hbar k/m)t'_{12}$).
After the scattering event the atom wave function evolves freely
until it reaches the second grating $G_2$.
It is important to note that the wave function $\psi_0^{\rm tr}(x,t)$,
associated with $\Delta k_x = 0$, describes also the evolution of
the wave behind the first grating when laser is off.

It is useful to parameterize wave function (\ref{e32}) in terms of
coordinate $y$ using the relation $\hbar t/m = y/k$,
\ba
 \fl \psi_{\Delta k_x}(x,t=my/\hbar k)
  & = & \frac{1}{\sqrt{2\pi}}\
  e^{i\Delta k_x(x + \Delta x_0) - i\Delta k_x^2 y/k} \nonumber \\
 \fl & & \qquad \times
   \int_{-\infty}^{\infty} c(k'_x) e^{-i{k'}_x^2 y/2k}
     e^{ik'_x (x + \Delta x_0 - \Delta k_x y/k)} dk'_x .
 \label{e32b}
\ea
The integrals in (\ref{e32}) and (\ref{e32b}) have no general
analytic solution, except for large $t$ or $y$ values.
In such a limit, when the dimensions of the diffracting object and the
wavelength of the diffracted beam are relatively small compared with
the typical propagation distances, the far-field or Fraunhofer
condition, $k{x'}^2/y \ll 1$ (with $x'$ being a measure of the
dimensions of the diffracting object), holds \cite{elmore} and
(\ref{e32b}) can be approximated (see \ref{appA}) by
\be
 \psi_0^{\rm tr}(x,t=my/\hbar k) = \sqrt{\frac{k}{2i\pi y}}\
  e^{ikx^2/2y} c(kx/y)
 \label{laseroff}
\ee
when the laser is off, and
\be
\fl
 \psi_{\Delta k_x}^{\rm tr}(x,t=my/\hbar k) =
  \sqrt{\frac{k}{2i\pi y}}\
  e^{ik(x + \Delta x_0)^2/2y - i\Delta k_x^2 y/2k}
   c[k(x + \Delta x_0)/y - \Delta k_x]
 \label{laseron}
\ee
for $\Delta k_x \in [0,2k_i]$ and the laser on.
By comparing (\ref{laseroff}) and (\ref{laseron}) we conclude that the
overall form of the atom probability density
$|\psi_{\Delta k_x}^{\rm tr}(x,t)|^2$ is the same as for
$|\psi_0^{\rm tr}(x,t)|^2$.
However, the former will display a shift or displacement along the
$x$-direction with respect to the latter given by
\be
 \Delta w_2 = \frac{\Delta k_x}{k}\ (y - y'_{12})
  = \left( \frac{\Delta k_x}{k} \right) y - \Delta x_0 .
 \label{e34}
\ee

The evolution of the wave function between $G_2$ and $G_3$ follows a
similar description to the one prior to the scattering event. Thus,
if the wave function incident onto $G_2$ is denoted as
$\psi_{{\rm inc},\Delta k_x}^{(2)}(x) \equiv
\psi_{\Delta k_x}^{\rm tr}(x,t=my_{12}^{-0}/\hbar k)$, which arises
from evaluating (\ref{laseron}) at $y=y_{12}^{-0}$, just before the
second grating, then wave function evolution behind the second grating
($y>y_{12}$) is given by
\ba
 \psi_{\Delta k_x}^{(2)}(x,t) & = & \frac{1}{\sqrt{2\pi}}\
  \int_{-\infty}^\infty c_{\Delta k_x}^{(2)}(k_x)
   e^{ik_x x - i\hbar k_x^2 t/2m} dk_x \nonumber \\
  & = & \frac{1}{\sqrt{2\pi}}\
   \int_{-\infty}^\infty C_{\Delta k_x}^{(2)}(k_x,t) e^{ik_x x} dk_x ,
 \label{e20bb}
\ea
where the relation between the time $t$ and $y$ is now $y - y_{12} = vt$
and the momentum probability density reads as
\be
 c_{\Delta k_x}^{(2)}(k_x) = \frac{1}{\sqrt{2\pi}}\
  \int_{-\infty}^\infty T(x) \psi_{{\rm inc},\Delta k_x}^{(2)}(x)
  e^{-ik_x x} dx .
 \label{e16bb}
\ee

From (\ref{e20bb}) and (\ref{e16bb}) one finds by numerical integration
that the probability density incident onto $G_3$ for a given value of
$\Delta k_x \in [0, 2k_i]$ oscillates with period $d_g$.
This oscillatory pattern (figure~3 in \cite{arsenovic1}) is of
finite width and its position along $x$-axis depends on $\Delta k_x$.
In other words, the oscillatory
pattern corresponding to $\Delta k_x \ne 0$ is shifted, with respect to
the oscillatory pattern when laser is off, by the quantity
\be
 \Delta w_3 = \frac{\Delta k_x}{k}\ (2y_{12} - y'_{12})
  = \left( \frac{\Delta k_x}{k} \right) 2y_{12} - \Delta x_0 ,
 \label{e37}
\ee
which arises after considering the shift of the wave function at $G_2$
(according to (\ref{e34})) and the influence of $\Delta k_x$ on the
propagation direction of the wave function emerging from $G_2$.
This estimate of $\Delta w_3$ is consistent with the shifts determined
through the numerical evaluation of the squared modulus of
$\psi_{\Delta k_z}^{(2)}(x,t=my_{23}/\hbar k)$
\cite{arsenovic1,arsenovic2}.


\subsection{Atomic flux behind the interferometer}
\label{sec32}

In order to compare the results obtained from the theoretical model
exposed above with the experimental data \cite{chapman}, we have first
considered the number of atoms transmitted through $G_3$ that undergo
a change of momentum $\Delta k_x$ during the scattering process.
This number is proportional to
\be
 \tilde{T}(y'_{12}, \Delta k_x, \Delta x_3) = \int_{\rm slits}
  \left\arrowvert \psi_{\Delta k_x}^{(2)} (x, t = my_{23}/\hbar k)
  \right\arrowvert^2 dx ,
 \label{e53}
\ee
where $\Delta x_3$ is a lateral shift of the third grating with
respect to the alignment of $G_2$ and the integration limits extend
over the region covered by the central maximum at $G_3$.
By numerical integration with the wave function determined as
described in the previous section, it has been found
\cite{arsenovic1,arsenovic2} that (\ref{e53}) the transmitted flux
(\ref{e53}) is a simple periodic function:
\be
 \tilde{T}(y'_{12}, \Delta k_x, \Delta x_3) = a
  + b \cos (2\pi \Delta x_3 /d_g + \Delta k_x d_p) ,
 \label{e54}
\ee
where $d_p$ is defined in (\ref{dp}), and $a$ and $b$ are constants
independent of $y'_{12}$ and $\Delta k_x$.
Far from the grating (i.e., large values of $y'_{12}$), the distance
$d_p$ is equal to the separation between the paths associated
with the zeroth and first order interference maxima of the atomic
wave diffracted by $G_1$ (see figure~\ref{fig1}a).
However, near the grating the emergent diffraction pattern is far more
complex than a series of well defined paths, obeying a Talbot-like
carpet structure \cite{sanz-talbot}.
This implies, as explained after (\ref{dist}) and in
\cite{arsenovic1} that $d_p$ should not be interpreted as the distance
between two atomic paths in the region covered by the laser light, for
in this region there are, actually, many more paths than simply two,
as it is generally assumed \cite{cronin1,chapman}.

The results reported in \cite{chapman} essentially come from two types
of measurements.
The first type consists of simply counting {\it all} atoms that pass
through $G_3$; in the second type, only a certain {\it subset} of the
transmitted atoms are counted or {\it postselected}, in particular
those with a certain momentum direction, which is done by positioning
an additional slit beyond $G_3$ (see figure~\ref{fig1}b).
Therefore, the observable is not $\tilde{T}(y'_{12},\Delta k_x,
\Delta x_3)$ in general, but its integral over a set of transferred
momenta $\Delta k_x$,
\ba
\fl
 T(y'_{12}, \Delta x_3) & = & \int_0^{2k_i} \tilde{P} (\Delta k_x)
  \tilde{T}(y'_{12}, \Delta k_x, \Delta x_3) d(\Delta k_x)
 \nonumber \\
 \fl
 & = & \int_0^{2k_i} \tilde{P} (\Delta k_x)
  \left[ a + b \cos (2\pi \Delta x_3/d_g + \Delta k_x d_p) \right]
   d(\Delta k_x) ,
 \label{e58}
\ea
where the weight $\tilde{P}(\Delta k_x)$ denotes the {\it transversal
momentum transfer distribution} of the detected atoms.
More specifically, this quantity is the product of the atom momentum
transfer distribution $P_0(\Delta k_x)$ and the
distribution function $P_s(\Delta k_x)$ characterizing the way
how the atoms are selected ({\it postselected}) by their momentum
beyond the interferometer.
That is, we have $\tilde{P}(\Delta k_x) = P_0(\Delta k_x)
P_s(\Delta k_x)$.
In particular, when the postselection process will be included, we
shall refer to the normalized $\tilde{P}$ function as the
{\it postselection momentum transfer distribution}.
Thus, if $P(\Delta k_x) \equiv \tilde{P}(\Delta k_x)/\Gamma$, with
$\Gamma \equiv \int_0^{2k_i} \tilde{P}(\Delta k_x)d(\Delta k_x)$,
is the corresponding normalized distribution, it is straightforward
to verify that (\ref{e58}) reads as
\be
 T(y'_{12}, \Delta x_3) = a
  + b \mathcal{V} \cos (2\pi \Delta x_3/d_g + \varphi ) ,
 \label{e62}
\ee
where the quantities $\mathcal{V}$ and $\varphi$ represent the
{\it fringe visibility} or {\it relative contrast} and the
{\it phase-shift}, respectively, and are determined through
the relations
\be
 \mathcal{V} \equiv \sqrt{I_r^2 + I_i^2} , \qquad
 \tan \varphi \equiv \frac{I_i}{I_r} ,
\ee
with
\be
 \begin{array}{rcl}
 I_r & \equiv & \int_0^{2k_i} P(\Delta k_x) \cos (\Delta k_x d_p)
  d(\Delta k_x) , \\ & & \\
 I_i & \equiv & \int_0^{2k_i} P(\Delta k_x) \sin (\Delta k_x d_p)
  d(\Delta k_x) .
 \end{array}
 \label{group}
\ee
From a practical point of view, in order to evaluate $\mathcal{V}$ and
$\varphi$, it is useful to introduce the complex integral
\be
 I \equiv \int_0^{2k_i} P(\Delta k_x) e^{i\Delta k_x d_p}
  d(\Delta k_x) = I_r + i I_i ,
 \label{e65}
\ee
so that
\be
 \mathcal{V} = \sqrt{I \cdot I^*} , \qquad
  \varphi = - \frac{i}{2}\ \ln \left( \frac{I}{I^*} \right) .
\ee
Taking this into account together with the standard definition of
fringe contrast \cite{mandel2}, from (\ref{e62}) we find
\be
 \mathcal{C} =
  \frac{T_{\rm max} - T_{\rm min}}{T_{\rm max} + T_{\rm min}}
   = \frac{|b|}{a}\ \mathcal{V} .
 \label{e67}
\ee
When the laser is off, $\Delta k_x = 0$ and hence
$T(y'_{12},\Delta x_3) = \tilde{T}(y'_{12}, 0, \Delta x_3) =
a + b \cos (2\pi \Delta x_3/d_g)$ and $\mathcal{C}_0 = |b|/a$.
The relative contrast then reads as
\be
 \frac{\mathcal{C}}{\mathcal{C}_0} = \mathcal{V} ,
 \label{e70}
\ee
which is a function of the ratio $d_p/\lambda_i$ ($\lambda_i$ is the
scattering photon wavelength), as it will be seen below.


\section{Numerical results}
\label{sec4}

In order to compare with the experiment, below we present some
calculations, where
we have considered the same parameter values used in
the experiment \cite{chapman}:
$v = 1400$~ms$^{-1}$, $k = m_{\rm Na} v/\hbar =
5.09067$$\times$10$^{11}$~m$^{-1}$, $\lambda_i = 589$~nm ($k_i =
1.06675$$\times$10$^7$~m$^{-1}$), $y_{12} = y_{23} = 0.65$~m,
$d_g = 2$$\times$10$^{-7}$~m and $\delta = 1$$\times$10$^{-7}$~m.
To evaluate the wave function, we have considered a total number of
illuminated slits $n = 24$ in $G_1$, which is an acceptable range
compared with experimental atomic beam cross-sections
(i.e., the coherence length of the atoms arriving in the grating)
\cite{turchette}.

\begin{figure}
 \begin{center}
 \epsfxsize=13cm {\epsfbox{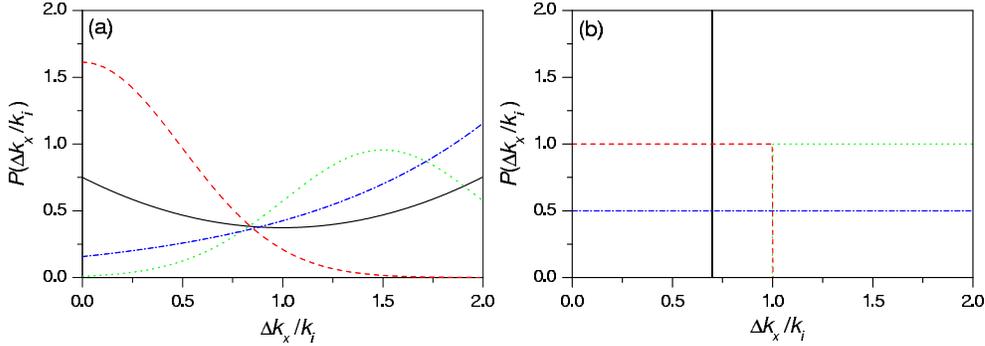}}
 \caption{\label{fig2}
  Transversal momentum transfer distributions as a function of the
  ratio between the transferred momentum and the incident photon wave
  number, $\Delta k_x/k_i$.
  In panel (a): bare momentum transfer distribution $P_0$ (black solid
  line) and postselection momentum transfer distributions $P_{\rm I}$
  (red dashed line), $P_{\rm II}$ (green dotted line) and $P_{\rm III}$
  (blue dashed-dotted line), as considered in the experiment
  \cite{chapman} (the colors follow those of figure~\ref{fig1}b).
  In panel (b): theoretical momentum transfer distributions $P_\delta$
  (with $k_\delta = 0.7k_i$; vertical black solid line), $P_1$ (red
  dashed line), $P_2$ (green dotted line) and $P_c$ (blue dashed-dotted
  line).
  All curves are normalized to unity within the interval $0 \le \Delta
  k_x/k_i \le 2$.
  See text for particular details on the values of the parameter.}
 \end{center}
\end{figure}

Apart from the Mandel distribution \cite{mandel1}, which accounts for
the bare transversal momentum transfer distribution, to compare with
the experiment we have also considered the three postselection momentum
transfer distributions used in the experiment, denoted by $P_{\rm I}$,
$P_{\rm II}$ and $P_{\rm III}$).
These distributions correspond to the combined effect of the momentum
transfer process (described by Mandel's distribution) and
three different particular selections (postselections of atomic momenta
(each one given by a different $P_s$ distribution), which are achieved
by arranging a slit behind $G_3$ in three different positions (see
figure~\ref{fig1}b).
The dependence of these four momentum transfer distributions as a
function of the ratio between the transferred momentum and the incident
photon wave number, $\Delta k_x/k_i$, is displayed in
figure~\ref{fig2}a.
Apart from these distributions, we have also considered several other
theoretical forms for the momentum transfer distribution of the
detected atoms, which are of interest to further analyze and better
understand the dependence of coherence and visibility on the
experimental distributions.
In particular, a Dirac $\delta$-function distribution ($P_\delta$) and
three constant distributions, $P_c$, $P_1$ and $P_2$, uniform over the
intervals $[0,2k_i]$, $[0,k_i]$ and $[k_i,2k_i]$, respectively.
These four distributions are displayed in figure~\ref{fig2}b.

\begin{figure}
 \begin{center}
 \epsfxsize=13cm {\epsfbox{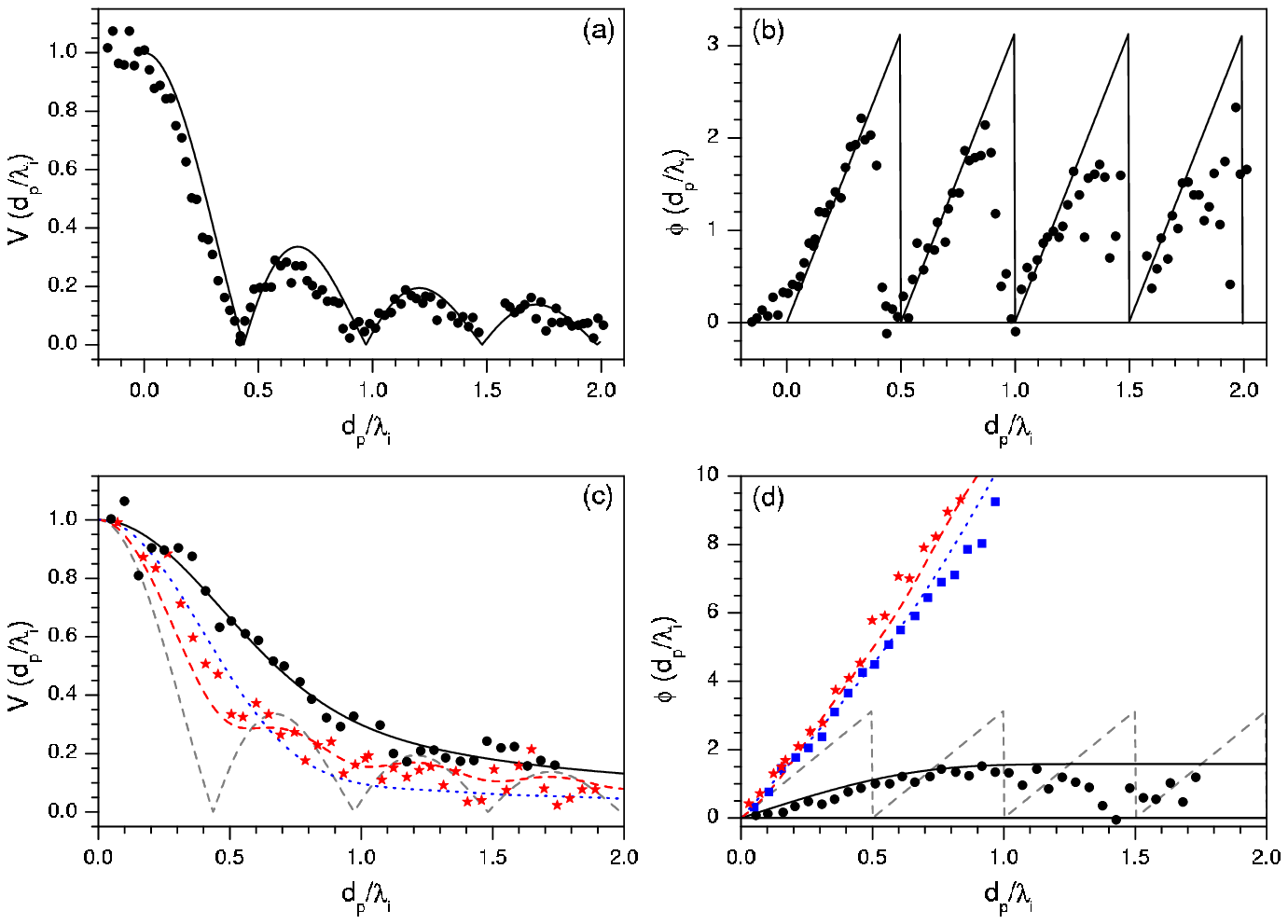}}
 \caption{\label{fig3}
  Functional dependence of the relative contrast (panels (a) and (c))
  and the phase shift (panels (b) and (d)) on the momentum transfer
  distributions displayed in figure~\ref{fig2}a, and as a function of
  the ratio $d_p/\lambda_i$.
  {\bf Top:} Theoretical curves (solid line) and experimental data
  (black solid circles) for the bare momentum transfer distribution
  $P_0$.
  {\bf Bottom:} Theoretical curves (lines) and experimental data
  (symbols) for the postselection momentum transfer distributions:
  $P_{\rm I}$ (black solid line/black solid circles), $P_{\rm II}$
  (blue dotted line/blue squares; no experimental data were available
  for the corresponding relative contrast) and $P_{\rm III}$
  (red dashed line/red stars).
  To compare with, the theoretical curves for the bare momentum
  transfer distribution $P_0$ have also been included, being denoted
  with the gray dashed line.
  The experimental data have been extracted from \cite{chapman};
  see text for particular details on the values of the parameters.}
 \end{center}
\end{figure}

A straightforward evaluation according to the method indicated
at the end of Section~\ref{sec32},
leads us to the following expressions for
the visibility and phase shift associated with these distributions:
\begin{enumerate}
\item As shown by Mandel \cite{mandel1}, for photons incident with a
momentum $k_i$, the transversal momentum transfer distribution can
be expressed as \cite{mandel2,mandel1}
\be
 P_0 (\Delta k_x) = \left( \frac{3}{8k_i} \right)
  \left[ 1 + \left( 1 - \frac{\Delta k_x}{k_i} \right)^2 \right] .
 \label{e56}
\ee
In this case, the visibility and phase shift read as
\ba
 \mathcal{V}_0 & = &
 \frac{3}{2} \frac{1}{k_i d_p} \left[
  \left( 1 - \frac{1}{k_i^2 d_p^2} \right)
    \sin (k_i d_p) + \frac{1}{k_i d_p} \cos (k_i d_p) \right] ,
 \label{e74} \\
 \varphi_0 & = & k_i d_p = \frac{2\pi d_p}{\lambda_i} ,
 \label{e75}
\ea
which are both functions of the ratio $d_p/\lambda_i$ (black solid
lines in figures~\ref{fig3}a and \ref{fig3}b).
As it can be seen, we find a good agreement between these
theoretical expressions and the experimental data (black solid circles)
without taking into account any fitting procedure.
Both the coherence losses and subsequent regains are thus accounted for
without abandoning the idea of pure state to describe the full
evolution of the atom.

\item The case of $P_{\rm I}$ is simulated by a half-Gaussian,
\be
 P_{\rm I} (\Delta k_x) = 2/Nk_i\sqrt{\pi}\ e^{- (\Delta k_x/N k_i)^2} ,
  \quad \Delta k_x \geq 0 ,
 \label{e71}
\ee
where $N$ determines the width of the Gaussian (here, we have chosen
$N = 0.7$, so that $P_{\rm I} (2k_i) \approx 0$).
In this case (see \ref{appA}),
\ba
 \mathcal{V}_{\rm I} & = &
 \frac{ |{\rm erf}(2/N - i\alpha) + {\rm erf}(i\alpha)|}
  {{\rm erf}(2/N)}\ e^{-\alpha^2/4} ,
 \label{eq-g13b} \\
 \varphi_{\rm I} & = & \frac{1}{2i}\ \ln \left[
   \frac{{\rm erf}(2/N - i\alpha) + {\rm erf}(i\alpha)}
    {{\rm erf}(2/N + i\alpha) + {\rm erf}(-i\alpha)} \right] ,
 \label{eq-g14b}
\ea
where $\alpha = Nk_id_p$.
As seen in figures~\ref{fig3}c and \ref{fig3}d (black solid lines),
there are no recurrences in $\mathcal{V}_{\rm I}$ (they are completely
damped), while $\varphi_{\rm I}$ approaches a constat value of $\pi/2$
as $d_p/\lambda_i$ increases.
Again, as it can be seen, we find a fair agreement with the
experiment (black solid circles).

If instead of $\eta = 0$, one would choose $\eta = 1$, i.e, the mirror
image of $P_{\rm I}$ with respect to $\Delta k_x = k_i$, then
\ba
 \mathcal{V}'_{\rm I} & = &
 \frac{ |{\rm erf}(2/N - i\alpha) + {\rm erf}(i\alpha)|}
  {{\rm erf}(2/N)}\ e^{-\alpha^2/4} ,
 \label{eq-g13bb} \\
 \varphi'_{\rm I} & = &  2k_i d_p + \frac{1}{2i}\ \ln \left[
   \frac{{\rm erf}(2/N + i\alpha) + {\rm erf}(-i\alpha)}
    {{\rm erf}(2/N - i\alpha) + {\rm erf}(i\alpha)} \right] .
 \label{eq-g14bb}
\ea
That is, the visibility is the same in both cases, but
$\varphi'_{\rm I} = 2k_i d_p - \varphi_{\rm I}$ is an increasing linear
function of $d_p/\lambda_i$ (after $\varphi_{\rm I}$ reaches its
maximum, steady value).

\item For $P_{\rm II}$ we consider a displaced Gaussian,
\be
 P_{\rm II} (\Delta k_x) =
  2/Nk_i\sqrt{\pi}[1 + {\rm erf}\ (1/2N)]
   \ e^{- [(\Delta k_x -3k_i/2)/N k_i]^2} ,
 \label{e72}
\ee
with its maximum at $\Delta k_x = 3k_i/2$ and $N = 0.7$, as before, so
that $P_{\rm II} (2k_i + 3k_i/2) \approx 0$.
With this, we find
\ba
 \mathcal{V}_{\rm II} & = &
 \frac{ |{\rm erf}(1/2N - i\alpha) + {\rm erf}(3/2N + i\alpha)|}
  {{\rm erf}(1/2N) + {\rm erf}(3/2N)}\ e^{-\alpha^2/4} ,
 \label{eq-g13bbb} \\
 \varphi_{\rm II} & = & \frac{3k_i d_p}{2} +
   \frac{1}{2i}\ \ln \left[
   \frac{{\rm erf}\ (1/2N - i\alpha) + {\rm erf}\ (3/2N + i\alpha)}
  {{\rm erf}\ (1/2N + i\alpha) + {\rm erf}\ (1/2N - i\alpha)} \right] ,
 \label{eq-g14bbb}
\ea
which are represented by blue dotted lines in figures~\ref{fig3}c and
\ref{fig3}d.
In this case, since there relative contrast is very similar to
that found for $P_{\rm I}$, no experimental data were reported.
We only have experimental results for the phase shift (blue squares in
figure~\ref{fig3}d), where a good agreement is also found.

\item $P_{\rm III}$ is described by means of an increasing exponential,
\be
 P_{\rm III} (\Delta k_x) = \epsilon/k_i (1 - e^{-2\epsilon})\
  e^{\epsilon (\Delta k_x/k_i - 2)} ,
 \label{e73}
\ee
where $\epsilon = 1$ is the increase rate (see blue dashed-dotted line
in figure~\ref{fig2}a).
This distribution leads to
\ba
 \mathcal{V}_{\rm III} & = & \frac{\epsilon}{1 - e^{-2\epsilon}}
 \frac{\sqrt{1 + e^{-4\epsilon} - 2 e^{-2\epsilon} \cos (2k_i d_p)}}
  {\sqrt{\epsilon^2 + (k_i d_p)^2}} ,
 \label{e83} \\
 \varphi_{\rm III} & = & (\tan)^{-1}
  \left\{ \frac{\sin (2k_i d_p - \phi) - e^{-2\epsilon} \sin \phi }
   {\cos (2k_i d_p - \phi) - e^{-2\epsilon} \cos \phi } \right\} ,
 \label{e84}
\ea
where $\phi = (\tan)^{-1} (k_i d_p/\epsilon)$.
As seen in figures~\ref{fig3}c and \ref{fig3}d (red dashed lines), now
$\mathcal{V}$ presents some damped recurrences and there is a
significant phase shift.
The same trend is also observed in the experimental data (red
stars), which follow very closely the behavior of the theoretically
predicted curves.
\end{enumerate}

There are several simple cases of particular interest,
because {\it grosso modo} they capture the essential features of the
distributions used in the experiment, which are the finite, uniform
momentum transfer distribution within the interval
$[k_1, k_2] \subset [0,2k_i]$, being zero everywhere else,
\be
 P_u (\Delta k_x) = \frac{1}{k_2 - k_1} ,
 \label{eq-44}
\ee
for $\Delta k_x \in [k_1, k_2]$.
For this form we find
\ba
 \mathcal{V}_u & = & \left\arrowvert
 {\rm sinc} \left[ \frac{(k_2 - k_1)d_p}{2} \right] \right\arrowvert ,
 \label{eq-46} \\
 \varphi_u & = & \frac{(k_2 + k_1)d_p}{2} .
 \label{eq-47}
\ea
As can be noticed, the visibility is given in terms of the half
distance between the limits of the interval, $(k_2 - k_1)/2$, while
the phase-shift is proportional to their half sum, $(k_2 + k_1)/2$,
which corresponds to the average momentum. This implies that the
visibility will decay and oscillate faster as both $k_1$ and $k_2$
approach the limits of the interval, the phase behaving in a similar
manner (i.e., increasing). On the contrary, if $k_1 \to k_2$, we
will be approaching the limit described by $P_\delta$:
$\mathcal{V}_u$ will oscillate more and more slowly (behaving almost
constant up to very large values of $d_p/\lambda_i$), while its
phase will approach $k_2 d_p$.
Now we will analyze each one of these cases separately:
\begin{enumerate}[(a)]
\item For $P_\delta (\Delta k_x) = \delta(\Delta k_x - k_\delta)$ the
visibility is constant and equal to unity along the interval $[0,2k_i]$
(see black solid line in figure~\ref{fig4}a).
This means that a monochromatic event does not destroy the coherence of
the atom wave function, but it only produces a phase shift
$\varphi_\delta = k_\delta d_p$ (see figure~\ref{fig3}b).

\begin{figure}
 \begin{center}
 \epsfxsize=13cm {\epsfbox{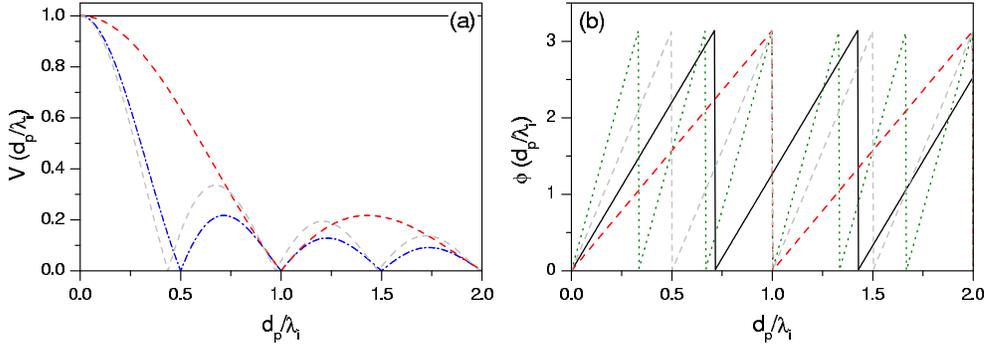}}
 \caption{\label{fig4}
  Functional dependence of the relative contrast (a) and the phase
  shift (b) on the momentum transfer distributions displayed in
  figure~\ref{fig2}b:
  $P_\delta$ (black solid line), $P_1$ (red dashed line), $P_2$ (green
  dotted line; $\mathcal{V}_2 = \mathcal{V}_1$ and no line can be seen)
  and $P_c$ (blue dash-dotted line).
  To compare with, the theoretical curves for the bare momentum
  transfer distribution $P_0$ have also been included, being denoted
  with the gray dashed line.
  See text for particular details on the values of the parameter.}
 \end{center}
\end{figure}

\item In the case $k_1 = 0$ and $k_2 = 2k_i$, $P_c (\Delta k_x) =
1/2k_i$, which is a rough approximation to $P_0$.
Here, we find
\ba
 \mathcal{V}_c & = & \frac{|\sin (k_i d_p)|}{k_i d_p} ,
 \label{e86} \\
 \varphi_c & = & k_i d_p .
 \label{e87}
\ea

\item If $k_1 = 0$ and $k_2 = k_i$, we have $P_1 (\Delta k_x) = 1/k_i$,
which roughly describes $P_{\rm I}$ and renders
\ba
 \mathcal{V}_1 & = & \frac{|\sin (k_i d_p/2)|}{k_i d_p/2} ,
 \label{e88} \\
 \varphi_1 & = & \frac{k_i d_p}{2} .
 \label{e89}
\ea

\item And, $k_1 = k_i$ and $k_2 = 2k_i$, we have $P_2 (\Delta k_x) =
1/k_i$, which can be an approximation to either $P'_{\rm I}$,
$P_{\rm II}$ or $P_{\rm III}$, and gives rise to
\ba
 \mathcal{V}_2 & = & \frac{|\sin (k_i d_p/2)|}{k_i d_p/2} ,
 \label{e90} \\
 \varphi_2 & = & \frac{3k_i d_p}{2} .
 \label{e91}
\ea
Notice that in this case and the previous one, the visibility is the
same, but not the phase shifts, which increases three times faster for
$P_2$ than for $P_1$.
\end{enumerate}
As it can be noticed, the functional forms found with our model
for the visibility and the phase shift associated with the different
momentum transfer distributions are in good agreement
with those reported in \cite{chapman}.

As it can be noticed, $\mathcal{V}_c$ vanishes for $d_p/\lambda_i=n/2$,
with $n$ being an integer, while $\mathcal{V}_1$ and $\mathcal{V}_2$
vanish when $d_p/\lambda_i = n$.
This is related to the fact that, for these three distributions, the
integrand in (\ref{group}) is a periodic function of $\Delta k_x$,
with period $2\pi/d_p$.
For $P_c$ the integration in (\ref{group}) is carried out over the
interval $[0,2k_i]$, which contains an integer number of periods when
$d_p/\lambda_i = n/2$.
For $P_1$ and $P_2$ the integration is performed over the intervals
$[0,k_i]$ and $[k_i,2k_i]$, respectively, which contain an integer
number of periods when $d_p/\lambda_i = n$.
Nevertheless, it is worth going further and analyzing the physical
reasons why the zeros of $\mathcal{V}_c$, $\mathcal{V}_1$ and
$\mathcal{V}_2$ appear at these values of $d_p/\lambda_i$. To start
with, let us remember that the phase $\Delta k_x d_p$ that appears
in $\tilde{T}(y'_{12}, \Delta k_x, \Delta x_3)$ arises as a
consequence of the shift $\Delta w_3$ along the $x$-axis at $G_3$
displayed by the atom wave function after the change of atomic
transverse momentum due to photon-atom scattering. This shift, which
is explicitly given by (\ref{e37}), contains the term $\Delta x_0$.
The latter is of the order of the grating constant $d_g$, as can be
noticed if we define $\Delta k_x = \eta k_i$, with
$0 \le \eta \le 2$ for $P_c$, $0 \le \eta \le 1$ for $P_1$, and
$1 \le \eta \le 2$ for $P_2$.
Thus, taking into account explicitly the value of $d_p$, we find
$\Delta x_0 = (d_p/\lambda_i) \eta d_g$, which
implies $0 \le \Delta x_0 \le (d_p/\lambda_i) 2d_g$ for $P_c$,
$0 \le \Delta x_0 \le (d_p/\lambda_i) d_g$ for $P_1$, and
$(d_p/\lambda_i) d_g \le \Delta x_0 \le (d_p/\lambda_i)$ for $P_2$.
Therefore, when $d_p/\lambda_i = 0.5$, $\Delta x_0$ lies within the
intervals $[0,d_g]$, $[0,d_g/2]$ or $[d_g/2,d_g]$ depending on we
have $P_c$, $P_1$ or $P_2$, respectively. This is why in the case of
a uniform momentum transfer distribution along the interval
$[0,2k_i]$ the total number of detected atoms (\ref{e58}) does not
depend on the lateral shift $\Delta x_0$ at $G_3$ and the contrast
is zero. However, if the transferred momentum spans the interval
$[0,k_i]$, the displacement of the wave function spans half the
grating constant and, therefore, the number of detected atoms will
depend on the lateral shift at $G_3$, then the contrast being
greater than zero. On the other hand, when $d_p/\lambda_i = 1$,
$\Delta x_0$ lies within the intervals $[0,2d_g]$, $[0,d_g]$ and
$[d_g,2d_g]$ for $P_c$, $P_1$ and $P_2$, respectively. In the three
cases the displacements thus span an integer number of grating periods.
Therefore, in any of these cases, the total number of detected atoms
will not depend on the lateral shift at $G_3$ and the contrast will
vanish (see figures~\ref{fig4}a and \ref{fig4}b).

It is insightful to analyze the experimental outcomes in the light of
the constant distributions.
One could therefore state that the contrast regain found in the
experiment, compared with the Mandel distribution, arises from the change
of the momentum transfer distribution of the detected atoms, which
is an objective effect.
Furthermore, the loss and revival of coherence in the case of the
Mandel distribution are also objective effects, which are related to
the properties of the atomic wave function incident onto $G_3$.


\section{Conclusions}
\label{sec5}

In spite of the details involved in entanglement-based models
aimed at describing
complementarity in experiments like the one here analyzed,
appealing to simpler models is also of interest in order
to understand the underlying physics, even if they are not fully
complete.
In the case dealt with here, we have considered a description based on
the recoil of the wave function describing the diffracted beam when a
photon impinges on it within the interferometer.
This model not only allows us to obtain a nice description of the
evolution of the wave function throughout the matter-wave Mach-Zehnder
interferometer, but also to explain the losses (e.g., the total loss at
$d_p = 0.5\lambda_i$), subsequent revivals (for $d_p/\lambda_i > 0.5$)
and regains (for all values of $d_p$ of experimental interest)
undergone by the (atom) fringe contrast in a very simple manner.
In particular, here we have presented how such effects arise when the
outgoing atomic probability density is sampled by a certain
momentum distribution, either Mandel's bare momentum transfer
distribution or the corresponding postselection ones.
In other words, these three effects can be attributed to the smearing
out of the interference pattern induced by the distribution of
transverse momentum that the photon or the postselection process
cause on the atomic beam.

In order to obtain some extra information, other momentum transfer
distributions of theoretical interest have also been considered.
In this regard, it was shown that, if the atoms
passing through $G_3$ could be selected
in such a way that only those with a chosen
value of transferred momentum would be detected, the contrast measured
would be constant, i.e., independent of $d_p/\lambda_i$ (see
figure~\ref{fig4}a for $P_\delta$).
On the contrary, if the statistical momentum distribution is constant
along the interval $[0, 2k_i]$, the interference contrast will be a
simple periodic function of $d_p/\lambda_i$ (see figure~\ref{fig4}a
for $P_c$).
These distributions allow us to understand the more complex situations
that takes place in real experiments, where the momentum transfer
distribution is given by the Mandel distribution.
In this case, in the light of the results obtained from the theoretical
momentum transfer distribution (in particular for $P_c$,
which is roughly similar; see figure~\ref{fig2}a and \ref{fig2}b), we
find how the losses and regains with $d_p/\lambda_i$ are associated
with the symmetry of this function with respect to $\Delta k_x = k_i$
(compare the gray curve for $P_0$ with the blue dashed-dotted one for
$P_c$ in figure~\ref{fig4}a).

We would like to stress that the
conclusions here obtained are also in agreement with those found
from postselection experiments \cite{rauch3,schleich} in neutron
interferometry \cite{rauch2,rauch3,rauch4}.
In this case, interference and coherence phenomena can be completely
hidden due to general averaging effects, but they can be recovered
even behind the interferometer if a proper postselection measurement
procedure is used.
This indicates that interference in phase space has to be considered
\cite{schleich} rather than the simple wave function overlap criterion
described by the coherence function.


\ack

MD, MB and DA acknowledge support from the Ministry of Science of
Serbia under Projects OI171005, OI171028 and III45016.
ASS acknowledges support from the Ministerio de Econom{\'\i}a
y Competitividad (Spain) under Projects FIS2010-22082 and
FIS2010-29596-C02-01, as well as for a ``Ram\'on y Cajal''
Research Fellowship.


\appendix


\section{}
\label{appA}

The approximations (\ref{laseroff}) and (\ref{laseron}) in the far
field have been obtained trough the following series of transformations
\cite{arsenovic1,davidovic}.
First, the wave function is expressed in terms of the initial wave
function behind the grating, which is done by substituting (\ref{e24})
and (\ref{e16}) into (\ref{e32b}),
\ba
 \fl
 \psi_{\Delta k_x}^{\rm tr}(x,y)
  & = & \frac{1}{\sqrt{2\pi}}\
  e^{i\Delta k_x(x + \Delta x_0) - i\Delta k_x^2 y/k} \nonumber \\
  \fl & & \times
  \int_{-\infty}^\infty dk'_x\ \frac{1}{\sqrt{2\pi}}
  \int_{-\infty}^\infty dx' \psi^{\rm tr}(x',0^+)
   e^{-i{k'}_x x'}
   e^{-i{k'}_x^2 y/2k}
   e^{ik'_x (x + \Delta x_0 - \Delta k_x y/k)} , \nonumber \\ \fl & &
 \label{e32b-app}
\ea
keeping in mind that the linear relation $t = my/\hbar k$ between $t$
and $y$ always holds.
Next, the integration over $k'_x$ in (\ref{e32b-app}) is carried out
taking into account the integral \cite{davidovic2}
\be
 \int_{-\infty}^\infty e^{-ux^2-vx} dx =
  \sqrt{\frac{\pi}{u}}\ e^{v^2/4u} ,
\ee
if ${\rm Re}(u) > 0$, ${\rm Re}(v) > 0$ or ${\rm Re}(u) = 0$,
${\rm Im}(u) \ne 0$ and ${\rm Re}(v) = 0$, ${\rm Im}(v) \ne 0$.
In doing so, we obtain the result
\ba
\fl
 \psi_{\Delta k_x}^{\rm tr}(x,y)
  & = & \frac{1}{2\pi}\
  e^{i\Delta k_x(x + \Delta x_0) - i\Delta k_x^2 y/k} \nonumber \\
  \fl & \fl & \fl \qquad \qquad \times \int_{-\infty}^\infty dx' \psi(x',0^+)
   \sqrt{\frac{k}{iy}}\ \sqrt{2\pi}\
   e^{i[k(x - x' + \Delta x_0) - \Delta k_x y]^2/2ky} .
 \label{e32b-appb}
\ea
In the far field approximation, the quadratic terms, ${x'}^2$, in the
exponent under the integral can be neglected, which yields
\ba
\fl
 \psi_{\Delta k_x}^{\rm tr}(x,y)
 & = & \frac{1}{\sqrt{2\pi}} \sqrt{\frac{k}{y}}\
  e^{-i\pi/4 + i\Delta k_x(x + \Delta x_0) - i\Delta k_x^2 y/k}
  e^{i[k(x + \Delta x_0) - \Delta k_x y]^2/2ky} \nonumber \\
 & & \fl \qquad \qquad \qquad \qquad \times \int_{-\infty}^\infty dx' \psi(x',0^+)
   e^{i[k(x + \Delta x_0) - \Delta k_x y]x'/y} .
 \label{e32b-appc}
\ea
After recognizing in the latter equation the expression from
(\ref{e16}), we find the form (\ref{laseron}) of the wave function
valid in the far field,
\be
\fl
 \psi_{\Delta k_x}^{\rm tr}(x,y) = \sqrt{\frac{k}{iy}}\
  e^{ik(x + \Delta x_0)^2/2y - i\Delta k_x^2 y/2k}
   c[k(x + \Delta x_0)/y - \Delta k_x] .
\ee


\section{}
\label{appB}

The analysis of Gaussian-shaped distributions (e.g., $P_{\rm I}$ and
$P_{\rm II}$) can be tackled in a general fashion as follows.
Consider the distribution is centered at $k_g=\eta k_i$, such that
$0 \le \eta \le 2$, i.e.,
\be
 P_g (\Delta k_x) = \gamma_g e^{- [(\Delta k_x - k_g)/N k_i]^2} .
 \label{eq-g1}
\ee
Here $N$ is some constant determining the width of the distribution
and $\gamma_g$ is the normalizing prefactor,
\be
 \gamma_g = \frac{2}{\sqrt{\pi} N k_i}
   \left[ {\rm erf}(\phi_+) + {\rm erf}(\phi_-) \right]^{-1} ,
 \label{eq-g2}
\ee
with ${\rm erf}(z)$ being the error function and
\be
 \phi_+ = \frac{2k_i - k_g}{Nk_i} = \frac{2 - \eta}{N} , \qquad
 \phi_- = \frac{k_g}{Nk_i} = \frac{\eta}{N} .
 \label{eq-g4}
\ee
Taking into account (\ref{eq-g1}), the integral (\ref{e65}) can be
expressed as
\be
 I_g = \left[ \frac{{\rm erf}(u_+) + {\rm erf}(u_-)}
    {{\rm erf}(\phi_+) + {\rm erf}(\phi_-)} \right]
    e^{- \alpha^2/4 + i\eta k_i d_p} ,
 \label{eq-g10b}
\ee
where $\alpha = N k_i d_p$.
Notice in the numerator of (\ref{eq-g10b})
that the error functions are complex, since their arguments,
\be
 \begin{array}{rcl}
 u_+ & = & \displaystyle \frac{2k_i - k_g}{Nk_i} - \frac{i\alpha}{2}
     = \frac{2 - \eta}{N} - \frac{i\alpha}{2} , \\ & & \\
 u_- & = & \displaystyle \frac{k_g}{Nk_i} + \frac{i\alpha}{2}
     = \frac{\eta}{N} + \frac{i\alpha}{2} ,
 \end{array}
 \label{eq-g9}
\ee
are also complex numbers.
Therefore, they will satisfy the properties ${\rm erf}(-z) =
-{\rm erf}(z)$ and $\overline{{\rm erf}(z)} = {\rm erf}(\bar{z})$.
From (\ref{eq-g10b}), the visibility and phase shift induced by $P_g$
are
\ba
 \mathcal{V}_g & = &
 \frac{ |{\rm erf}\ (u_+) + {\rm erf}\ (u_-)|}
  {{\rm erf}\ (\phi_+) + {\rm erf}\ (\phi_-)}\ e^{- \alpha^2/4} ,
 \label{eq-g13} \\
 \varphi_g & = & \eta k_i d_p + \frac{1}{2i}\ \ln
  \left[ \frac{{\rm erf}\ (u_+) + {\rm erf}\ (u_-)}
    {{\rm erf}\ (\bar{u}_+) + {\rm erf}\ (\bar{u}_-)} \right] .
 \label{eq-g14}
\ea
These two expressions can be evaluated for the half-Gaussian and
displaced Gaussian distributions considered in Section~\ref{sec3} by
simply setting $\eta = 0$ or $\eta = 3/2$, respectively.


\end{document}